\documentclass[prl,amsmath,amssymb,twocolumn,floatfix]{revtex4-2}
\usepackage{graphicx}
\usepackage{color}

\begin{document}

\title[Next-nearest neighbor acoustic surface waves]{Acoustic surface modes on metasurfaces with embedded \\ next-nearest neighbor coupling}

\author{D.~B. Moore}
\email{dm680@exeter.ac.uk}

 \author{J.~R. Sambles}
 \author{A.~P. Hibbins}
 \author{T.~A. Starkey}
 \author{G.~J. Chaplain}
\affiliation{Centre for Metamaterial Research and Innovation, Department of Physics and Astronomy, University of Exeter, Exeter EX4 4QL, United Kingdom}


\date{\today} 

\begin{abstract}
We design, simulate, and experimentally characterize an acoustic metasurface comprising of a 1D array of open, sound-hard, cavities, modulated with beyond-nearest-neighbor (BNN) couplings in the form of additional connecting cavities embedded beneath the surface. The hidden complex structure is realized readily with additive manufacturing techniques (3D printing). The dispersive properties of the supported localized acoustic surface waves are influenced by competing power-flow channels provided by the BNN couplings, that generate extrema in the dispersion spectra within the first Brillouin zone. The structure supports negatively dispersing `backwards' waves that we experimentally verify. Such structures thereby provide a route to enhanced acoustic sensing by acoustic metasurfaces.

\end{abstract}

\maketitle
\section{I Introduction}
Dispersion engineering is integral to the design of acoustic metasurfaces, with many applications across sensing, acoustic antennas, acoustic lensing, and tailored sound absorption \cite{Lan2019,Naify2013a,Ma2019,Popa2013,Chaplain2020,Ward2022,Jankovic2021a}, in addition to passive beam steering with topological surfaces \cite{Lai2021,Lin2022,Laforge2019,Ungureanu2021}. The wider theme of engineering dispersive materials has been a primary focus of metamaterial research for over a decade \cite{Li2021a,Chen2020,Assouar2018,Haberman2016,Ma2016}. Typically, acoustic metasurfaces are structured planar geometries, tessellated in some periodic distribution that manipulates acoustic waves through the collective interaction of resonators. 
\begin{figure}[h!]
    \centering
    \includegraphics[width = 0.355\textwidth]{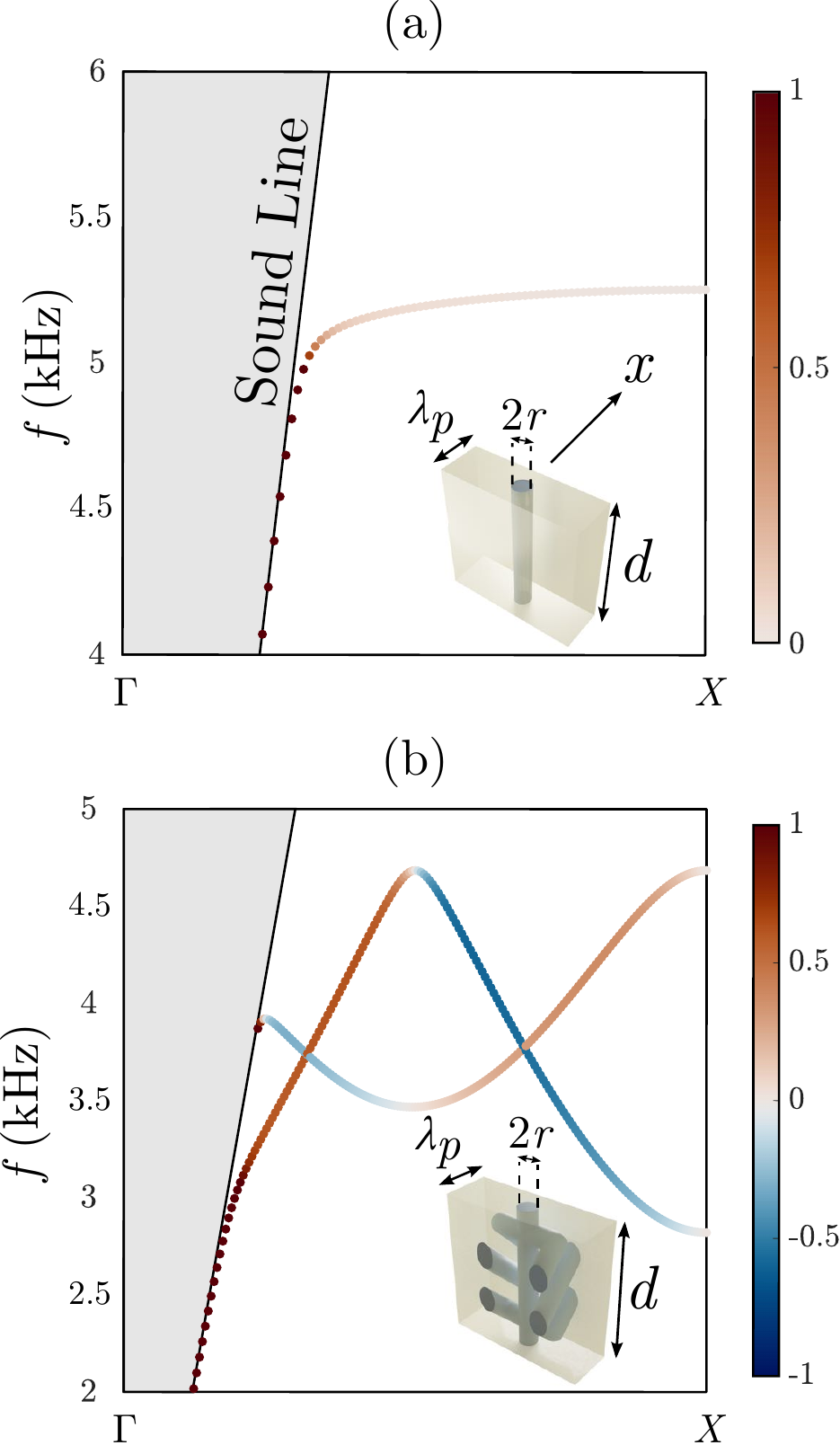}
    \caption{(\textbf{a-b}) FEM calculated dispersion relations (frequency against in-plane wave vector, $k_{||}$) showing the acoustic surface waves supported on a sample comprised of open hole cavities utilising NN coupling (\textbf{a}), and a revised surface containing BNN coupling channels within the substrate (\textbf{b}). Both samples have a periodicity $\lambda_p$ = 10 mm along the $x$-axis and are patterned with cylindrical resonators with depth $d$ = 30 mm and radius $r$ = 2 mm. The component of acoustic intensity ($I_{||}$) in the direction of propagation for each eigenmode is indicated by the color-map. The solid black line represents the sound-line ($\boldsymbol{k}_0$). Insets show example geometries with periodicity $\lambda_p$ in the propagation direction ($x$), depth $d$, and diameter $2r$.}
    \label{Fig_Intensity}
\end{figure}

The canonical example of a simple acoustic metasurface is a series of closely, periodically spaced resonant cavities; the resonators diffractively couple to their nearest neighbors (NN) through near-field effects, allowing an acoustic surface wave (ASW) to form, localized at the interface between the metasurface and the surrounding fluid. The formed ASW will propagate along the array of cavities, decaying evanescently away from the surface due to the effective impedance condition of the rigid cavities patterning the surface \cite{Kelders1998}. The substrate is assumed to be sound-hard, contributing no elastic motion. This example structure, with a single resonator per unit cell, supports only a single ASW associated with the cavity’s fundamental resonance (shown in Fig.~\ref{Fig_Intensity}(a)), and forms an acoustic analogue to electromagnetic spoof-surface plasmons on perfectly conducting patterned substrates \cite{pendry2004mimicking}. Increasing the number and complexity of the modes supported on the surface requires amending the structure factor, that is, the geometry and relative dimensions of the comprising resonant elements; the number of modes is linked to the degrees of freedom within the unit cell \cite{Cselyuszka2019,Moore2022,Hess2002}. An approach to engineering dispersion is to employ more complex structure factors that incorporate beyond-nearest-neighbor (BNN) coupling, via the introduction of additional resonators that acoustically connect a unit cell to its $L^{\text{th}}$ nearest-neighbor ($L > 1$), without altering the periodicity of the structure \cite{Chen2021,iglesias2021experimental,chaplain2022reconfigurable,Wang2022}. In a 1D NN-only coupled system, defined by a reciprocal lattice vector of size $2\pi/\lambda_p$, with $\lambda_p$ the periodicity, the typical dispersion is defined by modes that approach the 1\textsuperscript{st} Brillouin Zone Boundary (BZB) (where the in-plane wave vector, $k_\parallel = \pi / \lambda_p$) with corresponding vanishing group velocity, $v_{g_{||}} = \frac{\partial \omega}{\partial k_{||}}$; a standing wave is formed along the surface, satisfying the Bragg condition \cite{Ward2019}. In NN-only coupled systems there is therefore conventionally a single extremum in the dispersion curves when the mode reaches the asymptotic limit at the BZB. More complex structures that include symmetry broken geometries (either spatially, or in time) are some exceptions to this generalization \cite{chaplain2020delineating}. For systems with BNN coupling, it has been known for some 70 years that there will be $L - 1$ extrema within the 1\textsuperscript{st} Brillouin zone, causing regions in the dispersion curves to have alternating signs of the group velocity \cite{Brillouin1953,chaplain2022reconfigurable}, giving rise to modes that are said to have negative dispersion and are often termed ‘backwards waves'; the direction of wave propagation (momentum) and energy-flow are anti-parallel \cite{clarricoats1960non}.

Phenomena involving negative dispersion has been realized for some time with electromagnetic metasurfaces \cite{Tremain2018,Dockrey2016,Shin2006,Chen2018}, and recently BNN (sometimes termed `non-local' \cite{fleury2021non}) coupling in acoustic and elastic metamaterials has received renewed attention \cite{Chen2021,Martinez2021,Wang2022,chaplain2022reconfigurable}. Martinez \textit{et al.} \cite{Martinez2021} showed experimentally in both elastic and acoustic regimes that in a 3D structure comprised of BNN coupling to the $L = 3$ neighboring cell, a single fundamental mode is supported, exhibiting a maximum and minimum in its dispersion before the $1^{st}$ BZB. Here we extend these concepts to the canonical cavity acoustic metasurface by incorporating a hidden (i.e. embedded beneath the metasurface) next-nearest neighbor ($L=2$) coupling in the form of channels that reach-around the central open cavities. Nearest-neighbor coupling exists through near-field diffraction along the top and bottom of the surface, in addition to beyond-nearest neighbor coupling provided by the concealed connectors (shown in Fig.~\ref{Fig_Intensity}(b)). The additional structure provides an alternate route for power flow, and it is the competition between these two coupling paths that causes the extrema in the dispersion within the 1\textsuperscript{st} BZ. The result is regions in $k_\parallel$-space where there is a reversal in net power flow along the direction of propagation i.e. the surface supports negatively dispersing waves.  

Given the spatial symmetries associated with the unit cell, an anti-symmetric mode is supported in addition to the fundamental mode, see Fig.~\ref{Fig_Intensity}(c). The number of maxima can be tuned arbitrarily as a function of linking ${L}^{\text{th}}$ neighboring unit cells as well as breaks in the structures symmetry.

We demonstrate these effects experimentally for ASWs, using rapid prototyping (3D printing) to produce a scalable sample. This is an ever-increasingly popular method for realising complex, small-scale metamaterial geometries that would otherwise be difficult and expensive to manufacture \cite{gardiner2021additive}. In what follows we demonstrate that the observed negative dispersion is genuine in that it results from the competition of power flow rather than diffractive effects associated with, for example, glide symmetry \cite{Beadle2019,Moore2022}.

The structure of this paper is as follows. Firstly, in Section~\ref{sec:design}II, the metasurface geometry demonstrating BBN coupling is described and the supported eigenmodes evaluated, using Finite Element Method (FEM) numerical methods \cite{Comsol}. We introduce a second geometry, with a perturbed dispersion relation, by altering the boundary conditions along the metasurface and thereby breaking geometric symmetries associated with the structure. 3D-printing fabrication methods are then detailed, before the experimental results are presented in Section~\ref{sec:results}III.

\section{II Beyond Nearest Neighbors Coupling for Acoustic Surface Waves\label{sec:design}}

\begin{figure*}[t]
	\centering
		\includegraphics[width=0.8\textwidth]{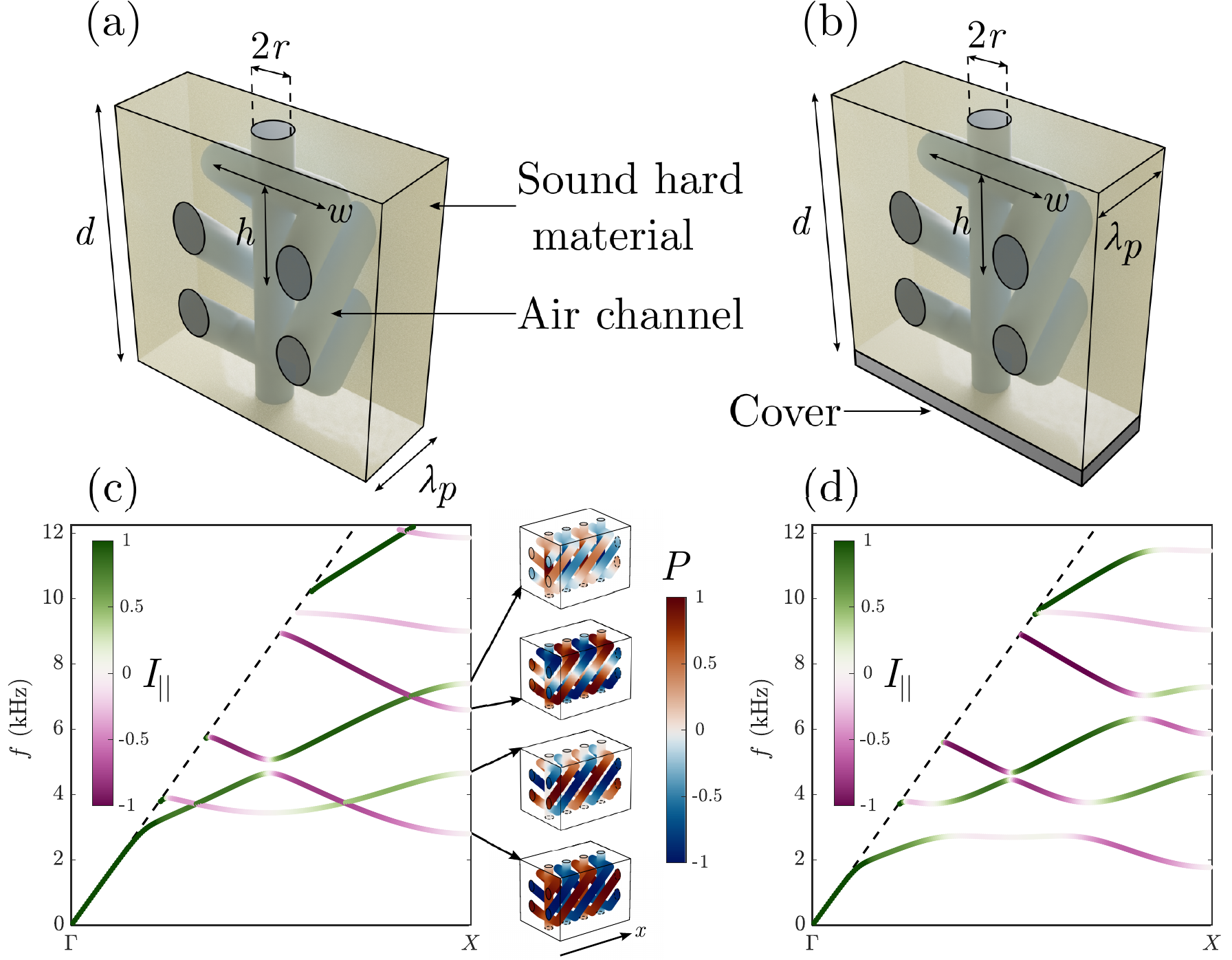}
	\caption{(\textbf{a,b}) Original and covered design unit cells. The sample has a periodicity $\lambda_p$ = 10 mm along the $x$ axis, and is patterned with cylindrical resonators with depth $d$ = 30 mm and radius $r$ = 2 mm. The internal coupling channels also have radius $r$, separated span-wise across the unit cell by width $w$ = 12.5 mm and vertically from the centre by height $h$ = 9.75 mm. \textbf{b} Illustrates the sound-hard boundary positioned along the covered side of the sample. (\textbf{c,d}) FEM calculated dispersion relations of eigenmode solutions showing the acoustic surface waves within the original (\textbf{c}) and covered (\textbf{d}) designs. The acoustic intensity ($I_{||}$, normalized to the maximum of the lowest band) in the direction of wave propagation for each eigenmode is indicated by the color-map. Instantaneous (normalized) pressure fields $P$ (with $ P_b=\left|{P-P_0}\right|$, where $P_0$ is the background pressure field) of eigenmodes taken at the 1\textsuperscript{st} BZB ($k_\parallel= X \equiv \pi / \lambda_p$) are shown for the original unit cell design (\textbf{a}).}
	\label{Fig_Modes}
\end{figure*}

\subsection{Design}
The design process from a nearest-neighbor-only acoustic metasurface, to one with additional BNN coupling. Figure~\ref{Fig_Intensity}(a) shows the simplest unit cell comprising an acoustic metasurface, a single periodic row of holes (cavities), of radius $r = 2$ mm, and depth $d = 30 $ mm, periodically spaced by $\lambda_p = 10 $ mm, coupled through NN near-field effects (diffractive coupling). The lowest frequency ASW supported by the structure is associated with the fundamental resonance of the structure (cavities) with limit frequency given approximately by $f_{res}=c/4(d+\Delta L)$, where $c$ represents the adiabatic speed of sound in air and $\Delta L$ represents the end correction for a pair of openings ($\Delta L = \frac{8r}{3 \pi} \approx 0.8 r$) \cite{Kinsler2000,Rayleigh1896a}. The in-plane wave vector of the surface mode, $k_\parallel$, is indistinguishable from the free-space wave or sound-line with $k_0=\omega/c$ for small values of $k_\parallel
$, for wave of radian frequency $\omega$. As $k_\parallel$ increases, the mode progressively disperses away from the sound-line, becoming strongly bound to the surface. The mode will disperse, with collinear phase and group velocity, towards its asymptotic limit until it meets the 1\textsuperscript{st} BZB with zero group velocity where it forms a standing wave see Fig. \ref{Fig_Intensity}(a).

\begin{figure}[h]
    \centering
    \includegraphics[width = 0.495\textwidth]{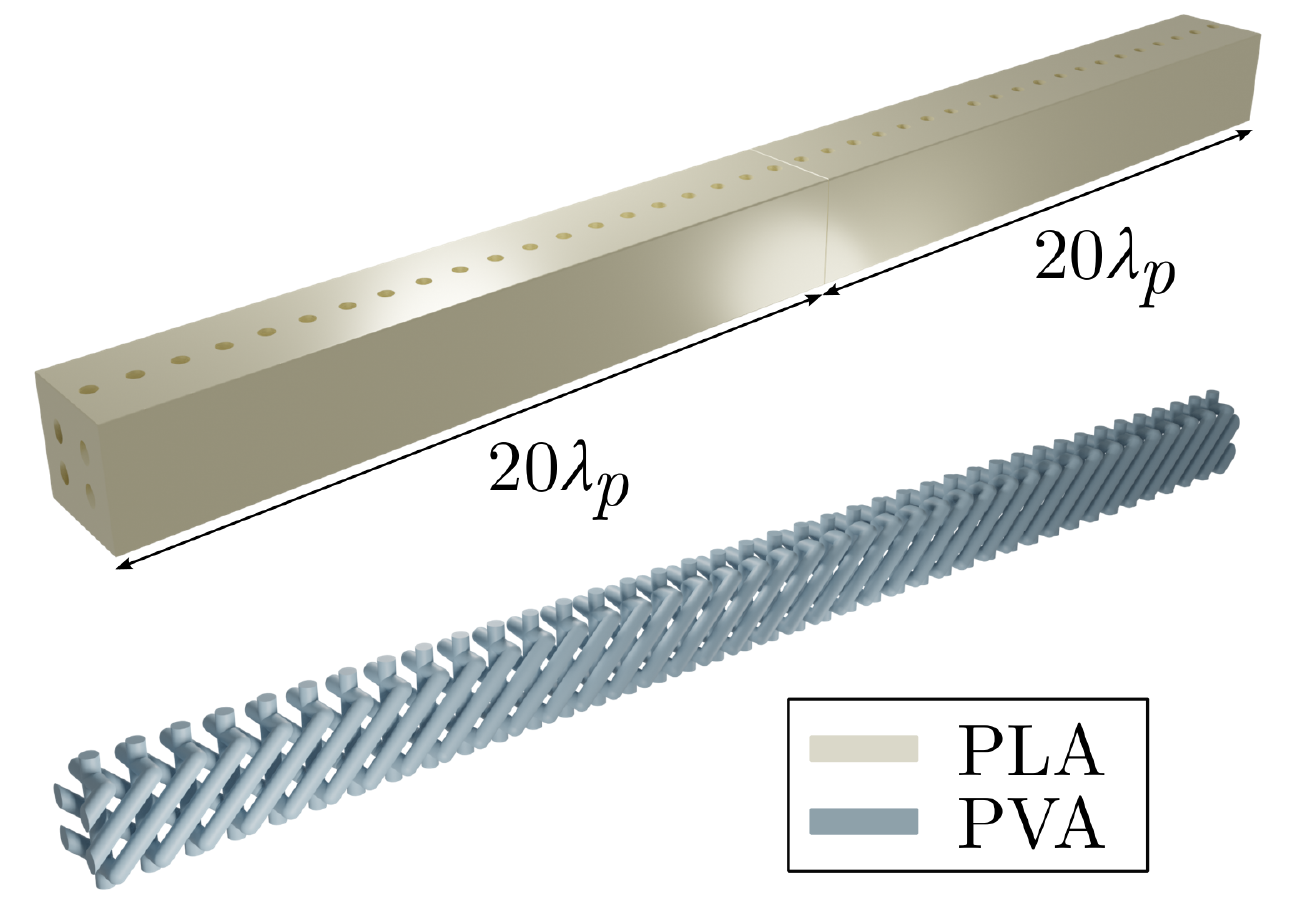}
    \caption{Rendered schematics of the sample, areas colored gold represent PLA, and grey represents water soluble PVA. The sample is comprised of two halves, each of 20 unit cells.}
    \label{Fig_Sample}
\end{figure}

To explore BNN coupling, the complexity of the structure factor needs to be increased. Increasing the number of resonators within the unit cell (thereby increasing the structure factor) allows additional modes to be supported by the surface \cite{Beadle2018} i.e. we are considering the complexity of the modes in terms of only the lowest order resonances, and not the infinite series supported by a single cavity. In this context, the complexity of such a structure is relatively low; all of the additional resonators are confined to the unit cell. Alternatively, cylindrical resonators (i.e. tubes within the structure) that link the $\pm~L^{th}$ nearest-neighbors, whilst bypassing the resonant elements comprising the cells between can be introduced, and hence provide a route to BNN coupling. The structure factor can be further modified through a rotational translation of the BNN resonators about the centre of the open cavity, producing the cross hatch structure shown in Fig. \ref{Fig_Modes}(a,b) and \ref{Fig_Sample}. The unit cell, at its lowest definable order of symmetry, now contains a resonator that passes through the cell, thereby only interacting acoustically with the other constituent resonators in the next-nearest cell \cite{Wigner1933,Slater1934,Brillouin1953}, see Fig. \ref{Fig_Modes}(a). The additional coupling channels have a fundamental resonance dictated by their length (typically lower in frequency than the open cavity due to their increased length). The BNN coupling manifests due to the overlap of the evanescent fields from these additional connectors with those of the open cavity.

The dispersion of the supported modes drastically changes from the case of a single resonant cavity, with extrema in the lowest order mode now existing within the first BZ, shown in Fig.~\ref{Fig_Modes}; a feature commonly associated with symmetry breaking or topological effects \cite{chaplain2020delineating,Ungureanu2021}. This implies that over certain regions of $k_\parallel$, the group velocity of the mode is negative, and the net power flow is travelling opposite to the direction of wave propagation (noted by the regions of negative intensity in Fig. \ref{Fig_Intensity}(b)); here, the integral of the intensity ($I_{||}$) is calculated along the direction of periodicity ($x$-axis) throughout the whole computational domain, providing the direction of the net flow of energy of the ASW. In the frequency domain the intensity is given by
\begin{equation}
    \boldsymbol{I} = \frac{1}{4}\left(p\boldsymbol{v}^{*} + p^{*}\boldsymbol{v} \right),
\end{equation}
where the velocity, $\boldsymbol{v}$ is related to the pressure $p$ through $\boldsymbol{v} = \frac{-1}{i\omega\rho}\nabla p$, with $*$ denoting complex conjugation.

Given the spatial symmetries present within the unit cell (where the connections between the coupling channels and open hole are mirrored about the centre of the unit cell), an additional anti-symmetric mode is also supported by the structure due to the coupling of adjacent resonators within the open cavity, see Fig. \ref{Fig_Modes}(c). The anti-symmetric mode is odd about the $xy$ plane passing through the centre of the structure and disperses with the opposite sign in $v_g$ and $I_{||}$ to the fundamental mode, with the modes exchanging sign of the net power flow when $k_\parallel = \frac{\pi}{L\lambda_p}$, see Fig. \ref{Fig_Modes}(c). The fundamental mode approaches the BZB beneath the anti-symmetric mode, as both the coupling channels within the unit cell are resonating in-phase relative to one another, thus having a lower frequency, see Fig. \ref{Fig_Modes}(c). For the anti-symmetric mode, at the band edge, both channels are resonating in anti-phase ($\pi$ radians difference) relative to one another, thereby halving the length of the resonator compared to that of the fundamental, hence increasing the resonant frequency, see Fig. \ref{Fig_Modes}(c). 

The propagation direction of the wave vector (either positive or negative) can be inferred from probing the near-field, since the evanescent fields of the ASW decay into the surrounding medium via the open cavities. The number of extrema in the dispersion before the 1\textsuperscript{st} BZB can be tuned by altering $L$ \cite{Brillouin1953}. In this study, the value of $L$ has been set to 2 (or next-nearest neighbor). Provided the required number of coupling channels can be contained within the unit cell while remaining acoustically isolated, any integer value for $L$ can be conceived. 

The symmetry of the unit cell can be broken by replacing one of the open ends of the open cavity with a sound-hard boundary condition, causing the fundamental and anti-symmetric modes to anti-cross, forming a bandgap, see Fig. \ref{Fig_Modes}(d). By covering one of the surfaces the coupling strength of the BNN channels closest to the closed end increases, biasing the power flow through that junction. The same number of extrema are present within the 1\textsuperscript{st} BZ. Additionally, similar results are found when reversing the direction of the BNN connections on one side of the unit cell, analogous to a mirror plane in the centre of the unit cell. 

\subsection{Fabrication: 3D printing}
\label{sec:fabrication}
The sample was 3D printed using an Ultimaker S5, extruding polylatic acid (PLA) filament. Support structures (for the inner tubes) were used to aid the printing process, created using a water soluble filament, polyvinyl alcohol (PVA) which was removed before measurements. The sample has length $L_{\text{total}} = 400$ mm, periodicity $\lambda_{p}$ = 10 mm along the $x$ axis (40 unit cells total), and is visibly patterned with cylindrical resonators with depth $d = 30$ mm and radius $r = 2$ mm, the coupling channels have the same radius, see Fig.~\ref{Fig_Modes}(a,b) for schematic and Fig.~\ref{Fig_Sample} for render. The sample was manufactured with a 40\% infill density to maintain a sound-hard boundary condition at the walls of the sample. The scale of the sample dimensions was carefully chosen to produce resonances optimally situated for detection experimentally; the frequency range of the sample can be tuned by scaling thegeometry dimensions. 

\section{III Experimental Results\label{sec:results}}
\begin{figure}[ht]
	\centering
		\includegraphics[width=0.4\textwidth]{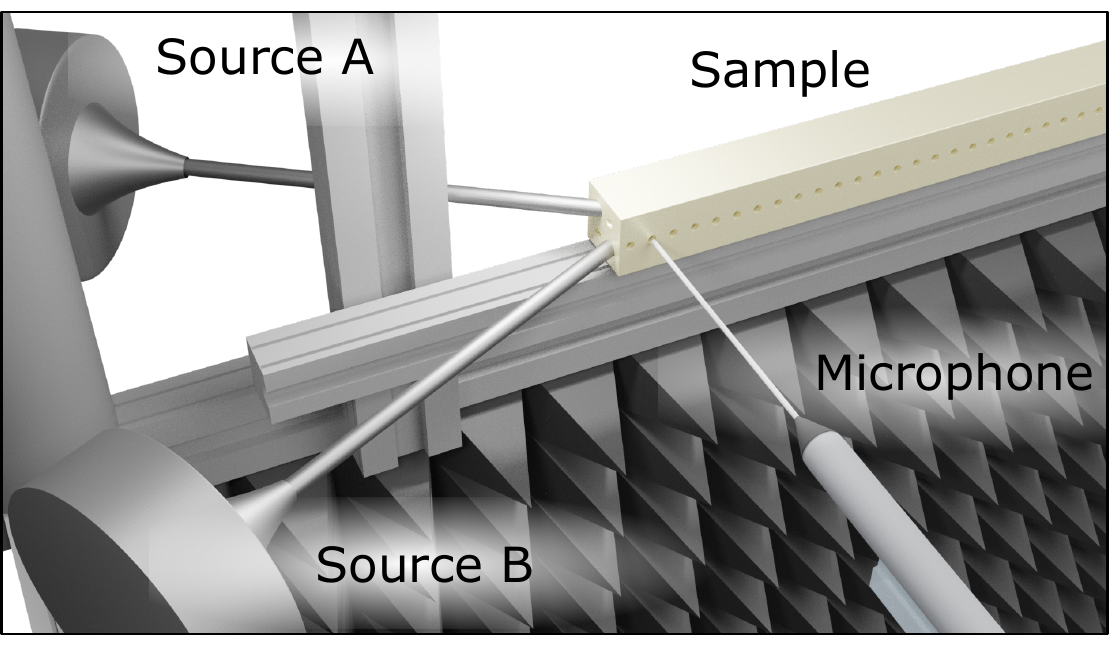}
	\caption{Render illustrating the acoustic measurement experiment, a pair of loudspeakers mounted within conical housings and attached to cylindrical waveguides are shown, positioned at the end of the sample and the microphone positioned over the open cavities.}
	\label{Fig_Lab}
\end{figure}

To verify the numerical predictions we experimentally characterized the ASW fields supported on the surface of the sample. Model dispersion relations as a function of the in-plane wavenumber, $k_{||}$, were obtained with COMSOL Multiphysics (version 5.6) \cite{Comsol} using the Finite Element Method. Dispersion relations shown are eigenmodes of a unit cell (see Fig.~\ref{Fig_Modes}(a,b)) with Floquet-periodic boundaries to represent an infinite sample. This model assumes the resonators are cylindrical and perforate/enclosed by an acoustically rigid surface. The dispersion for each excitation arrangement was characterized by measuring the acoustic near-field using a probe microphone mounted on a motorized \textit{xy} scanning stage. Samples were excited by a pair of Visaton SC5 13 mm tweeters mounted within conical attachments with 3 mm exit diameters. The exits of the conical housings were attached to cylindrical waveguides of length 150 mm, positioned within the coupling channels to allow the sources to be correctly positioned due to their size, see Fig. \ref{Fig_Lab}. The pair of loudspeakers could be driven in- or out-of-phase relative to one another. To minimize signal from direct radiation, the needle microphone (Br\"{u}el {\&} Kjær Probe Microphone type 4182) was positioned 0.5 mm over the centre of the open cavities. The surface was scanned at a resolution of 0.45 mm for a scan length of 400 mm ($x$-axis). A single cycle 5 kHz Gaussian envelope pulse (broadband) was used to excite the sample at each microphone position. An average was taken over eight measurements at each spatial position to improve the signal-to-noise ratio. The dispersion was characterized by performing a Fast Fourier Transform (FFT) on the measurement data to obtain the frequency (Fourier) amplitude of the real spatial component as a function of wave vector, $|k_{||}|$, thereby recovering the single-sided spectra whilst we only measure power flow away from the source towards the detector.

\begin{figure}[t]
	\centering
		\includegraphics[width=0.495\textwidth]{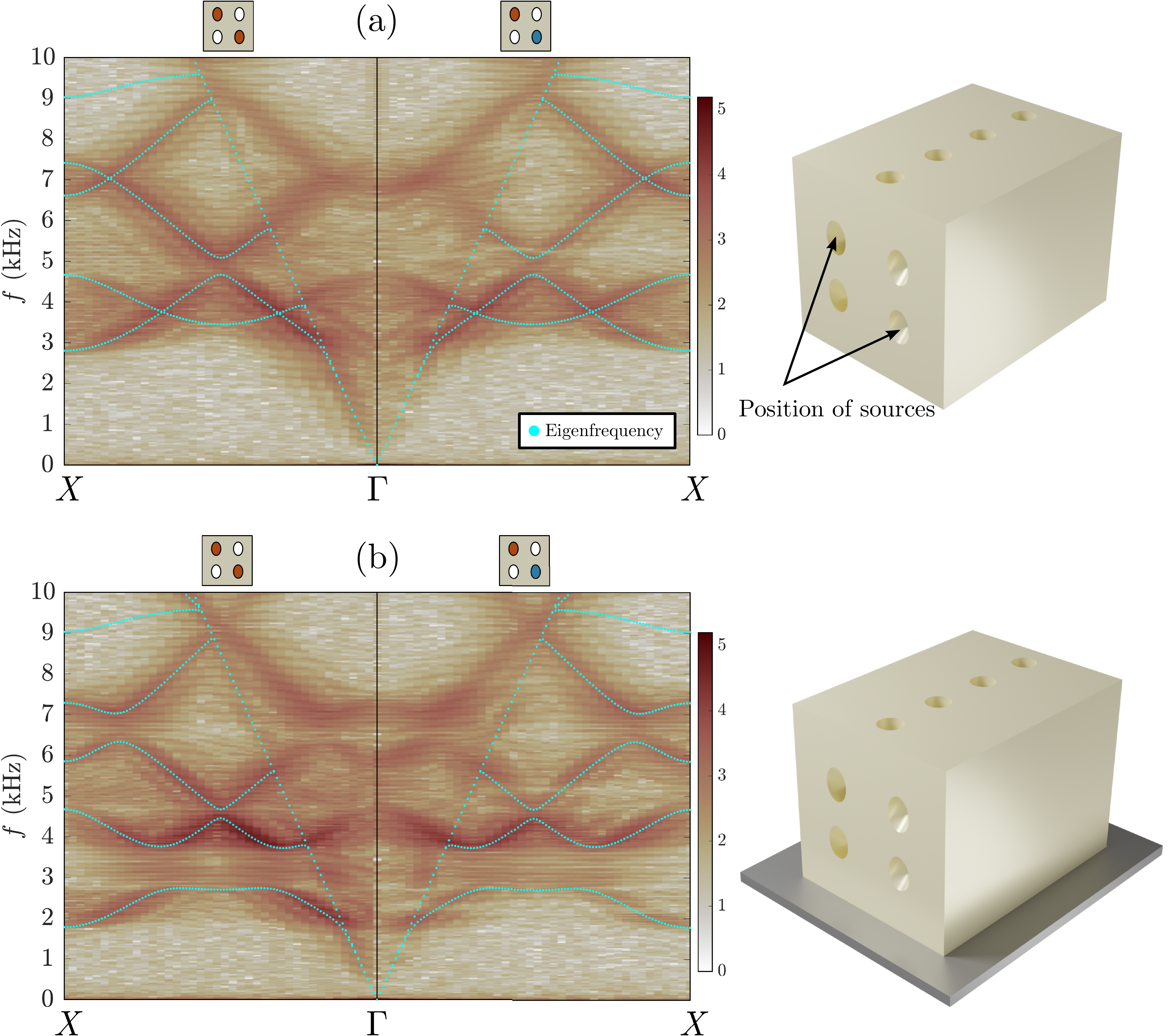}
    	\caption{Comparisons between FEM eigensolutions (cyan points) and logarithmic experimental Fourier spectra (color scale), obtained via the real FFT. Left column shows results for in-phase excitation (red points in schematic above the dispersion curves representing the source positions; right column for the out-of-phase excitation (red and blue points in schematic), for (\textbf{a}) original set-up (\textbf{b}) with one side covered. Note we plot $|k_{||}|$ on the $x$-axis.}
	\label{Fig_Disp}
\end{figure}

Figure~\ref{Fig_Disp} shows the measured dispersion of the modes of the sample, with and without altered symmetry, through the logarithm of the experimental Fourier amplitude (real FFT). Results from a loss-inclusive FEM model are overlaid as blue points, for both excitation field arrangements and each symmetry arrangement; with the left hand side of the dispersion relations from in-phase excitation of the sources and the right hand side showing the results from the out-of-phase excitation. Figure \ref{Fig_Modes}(c) shows the pressure field solutions for the anti-symmetric mode to have a zero pressure state between the opposing coupling channels ($\pi$ phase difference). The pressure must reach zero between the opposing resonators to satisfy the phase arrangement of the eigenmode. However, despite the zero pressure condition, the mode is detectable due to the pressure amplitude of the decaying fields above the surface. This is evidenced by the detected anti-symmetric modes in Fig.~\ref{Fig_Disp}. Analysing the pressure fields within the structure at the first extremum in the dispersion, the increased length of the coupling structures becomes apparent, as the pressure oscillations are confined to the coupling channels, while the free-space wavelength is still four times the lattice pitch. The resonance of the open cavities is not excited until the $\lambda/2$ condition of the substrate depth is satisfied. The fundamental mode leaves the sound-line with a near-constant group velocity; this frequency at which the mode leaves the sound-line, transitioning from the free-space speed of sound to that of a waveguide mode \cite{Beadle2019}. The reduction of the sound-speed in this region is reminiscent of coiled space metamaterials that provide a longer acoustic path length \cite{liang2012extreme}. Exciting the sample with a pair of monopole sources allows the phase and amplitude of the excitation field to match that of the mode. In Fig.~\ref{Fig_Disp}(b) the amplitude of the fundamental mode is decreased when the sample is excited with an excitation field which matches the anti-symmetric mode shape. 

Figure~\ref{Fig_Disp}(b) shows the measured dispersion of the sample with altered symmetry. Covering one side of the surface prevents energy from decaying into the surrounding air, increasing the amplitude through that junction, biasing the power flow. Replacing one of the open boundary conditions with a sound-hard condition alters the symmetry of the modes, lifting the degeneracy, forming a low-frequency bandgap across the 1\textsuperscript{st} BZ. The resonant condition of the cavity becomes approximately $\lambda/4$ due to the sound-hard boundary condition at the bottom of the open hole, decreasing the frequency of the fundamental resonance.

\section{IV Conclusions}
We have manufactured a sample using 3D printing which is visibly patterned with a 1D array of resonant cavities while embedding within the substrate, a network of coupling channels facilitating next-nearest-neighbor coupling. The eigenmodes supported by this surface have been calculated using numerical modeling and demonstrated experimentally. We have detected, through probing the near-field, the full dispersion spectra across the frequency range considered by matching the excitation symmetry to that of the supported eigenmodes. The dispersion of the supported modes is tuneable through manipulating the symmetry of the unit cell by covering select resonant cavities. Further frequency bands beyond those explored in this study can be achieved through scaling the geometry.

\section*{Data availability}
\noindent All data created during this research are openly available from the University of Exeter institutional repository at \href{https://ore.exeter.ac.uk/}{https://ore.exeter.ac.uk/}.

\section*{acknowledgements} The authors wish to acknowledge financial support from the Engineering and Physical Sciences Research Council (EPSRC) of the United Kingdom, via the EPSRC Centre for Doctoral Training in Metamaterials (Grant No. EP/ L015331/1). D.B.M., T.A.S., A.P.H. and J.R.S. acknowledge the financial support of DSTL. G.J.C. gratefully acknowledges financial support from the Royal Commission for the Exhibition of 1851 in the form of a Research Fellowship.

\section*{Competing interests}
\noindent The authors declare no competing interests.

\bibliography{library}

\end{document}